\documentclass[DIV=calc,paper=a4,fontsize=11pt,captions=tableheading]{scrartcl}
\usepackage[english]{babel}
\usepackage[babel]{csquotes}
\usepackage[T1]{fontenc}
\usepackage{times}
\usepackage[dvips]{graphicx}
\usepackage[a4paper,includehead,twoside,bindingoffset=2cm,scale=0.85]{geometry}
\usepackage[x11names,svgnames,table]{xcolor}
\usepackage[hang,small,labelfont=bf,up,textfont=it,up]{caption}
\usepackage{longtable} 
\usepackage{fancyhdr}
\usepackage{lastpage}
\usepackage{paralist} 

\usepackage[colorlinks=true,linkcolor=Firebrick4,urlcolor=SteelBlue4,citecolor=Firebrick4]{hyperref}

\title{Indexing Fe-phases in/on GaN using x-ray powder diffraction}
\author{Mauro Rovezzi}
\date{October 13, 2011}
\hypersetup{
  pdfkeywords={Fe-Nitrides, MOVPE growth},
  pdfsubject={Fe-Nitrides},
  pdfcreator={Mauro Rovezzi}}

\begin{document}

\maketitle

\setcounter{tocdepth}{3}

This document reports the x-ray powder diffraction main reflections (intensity threshold $\ge$ 100) for possible Fe-related phases forming during the metal-organic vapor phase epitaxy (MOVPE) growth of Fe in NH$_{3}$/H$_{2}$ mixture on wurtzite-GaN/sapphire. The 2$\theta$ values are given for Cu K$\alpha_{1}$ radiation (1.5406 \AA) in the range 25-100 deg (ordered by increasing 2$\theta$). The GaN(000\emph{l}) and Al$_{2}$O$_{3}$(000\emph{l}) are also reported for reference. These data are obtained from the Inorganic Crystal Structures Database (ICSD, \href{http://icsd.fiz-karlsruhe.de/}{http://icsd.fiz-karlsruhe.de/}). The access to ICSD has been provided by the European Synchrotron Radiation Facility (ESRF).

\rowcolors{2}{gray!35}{}
\begin{longtable}{|p{28mm}p{15mm}p{15mm}p{7mm}p{8mm}p{12mm}p{2mm}p{10mm}p{5mm}|}
\caption{Indexing Fe-phases in/on GaN using x-ray powder diffraction} \label{tab:Fephases}\\
\hline
 \rowcolor{gray!70} \textbf{Phase}       &  \textbf{Type}            &  \textbf{Spg.}   &  \emph{HKL}  &  2$\theta$  &  \emph{d}-sp.  &  \textbf{M.}  &  \textbf{Int.}  &  \textbf{Ref.}                \\
\hline
\endhead
\hline\multicolumn{9}{r}{Continued on next page}\
\endfoot
\endlastfoot
 $\epsilon$-Fe$_{2}$N                      &                           &  \verb~P312~     &         101  &    29.52  &        3.0238  &            6  &         1000.0  &  \cite{Jack:1952_AC}           \\
 $\epsilon$-Fe$_{3}$N                      &                           &  \verb~P312~     &         101  &    29.54  &        3.0212  &            6  &          629.3  &  \cite{Pinsker:1954_DAN}       \\
 Fe$_{24}$N$_{10}$                &                           &  \verb~P312~     &         210  &    29.59  &        3.0163  &            6  &          444.3  &  \cite{Jack:1952_AC}           \\
 Fe$_{24}$N$_{10}$                &                           &  \verb~P312~     &         120  &    29.59  &        3.0163  &            6  &          444.3  &  \cite{Jack:1952_AC}           \\
 $\epsilon$-Fe$_{3}$N                      &                           &  \verb~P312~     &         101  &    29.84  &        2.9915  &            6  &          641.1  &  \cite{Jack:1952_AC}           \\
 Fe$_{3}$O$_{4}     $                &  Al$_{2}$MgO$_{4}$          &  \verb~Fd-3mZ~   &         220  &    30.09  &        2.9678  &           12  &          284.0  &  \cite{Fleet:1981_ACB}         \\
 FeGa$_{3}$                  &  FeGa$_{3}     $      &  \verb~P42/mnm~  &         210  &    31.93  &        2.8008  &            8  &          202.7  &  \cite{Haussermann:2002_JSSC}  \\
 Fe$_{2}$O$_{3}     $                &  Al$_{2}$O$_{3}$    &  \verb~R-3c:H~   &         104  &    33.12  &        2.7028  &            6  &         1000.0  &  \cite{Blake:1966_AM}          \\
 Fe$_{24}$N$_{10}$                &  $             $        &  \verb~P312~     &         300  &    33.66  &        2.6601  &            6  &          135.5  &  \cite{Jack:1952_AC}           \\
 FeGa$_{3}        $                  &  FeGa$_{3}     $      &  \verb~P42/mnm~  &         112  &    34.00  &        2.6344  &            8  &          382.8  &  \cite{Haussermann:2002_JSSC}  \\
\hline
 \rowcolor{red!20}  GaN                  &  ZnS(2H)                  &  \verb~P63mc~    &         002  &    34.56  &        2.5931  &            2  &          360.5  &  \cite{Paszkowicz:2004_JAC}    \\
\hline
 Fe$_{3}$O$_{4}               $      &  Al$_{2}$MgO$_{4}$          &  \verb~Fd-3mZ~   &         311  &    35.44  &        2.5309  &           24  &         1000.0  &  \cite{Fleet:1981_ACB}         \\
 Fe$_{2}$O$_{3}               $      &  Al$_{2}$O$_{3}$    &  \verb~R-3c:H~   &         110  &    35.61  &        2.5190  &            6  &          732.7  &  \cite{Blake:1966_AM}          \\
 $\zeta$-Fe$_{2}$N $            $        &                           &  \verb~Pbcn~     &         021  &    37.46  &        2.3988  &            4  &          150.9  &  \cite{Hasegawa:2005_JAC}      \\
 $\epsilon$-Fe$_{3}$N $         $        &                           &  \verb~P312~     &         110  &    37.53  &        2.3945  &            6  &         1000.0  &  \cite{Pinsker:1954_DAN}       \\
 $\epsilon$-Fe$_{2}$N $         $        &                           &  \verb~P-3m1~    &         100  &    37.53  &        2.3946  &            6  &          275.5  &  \cite{Pinsker:1954_DAN}       \\
 $\epsilon$-Fe$_{2}$N $         $        &                           &  \verb~P312~     &         110  &    37.55  &        2.3935  &            6  &          497.2  &  \cite{Jack:1952_AC}           \\
 $\epsilon$-Fe$_{3}$N$_{1.39}   $      &  Mn$_{2}$N$_{1-x}$          &  \verb~P6322~    &         110  &    37.56  &        2.3928  &            6  &          156.1  &  \cite{Leineweber:2001_JAC}    \\
 $\epsilon$-Fe$_{2.3}$N $       $        &  NiAs $        $        &  \verb~P63/mmc~  &         100  &    37.64  &        2.3876  &            6  &         1000.0  &  \cite{Kim:994_JJAP}           \\
 Fe$_{3}$C                           &  Fe$_{3}$C            &  \verb~Pnma~     &         121  &    37.64  &        2.3878  &            8  &          267.9  &  \cite{Fruchart:1984_JSSC}     \\
 Fe$_{3}$C$                   $        &  Fe$_{3}$C$      $      &  \verb~Pnma~     &         210  &    37.74  &        2.3818  &            4  &          244.5  &  \cite{Fruchart:1984_JSSC}     \\
 $\epsilon$-Fe$_{3}$N $         $        &  Fe$_{3}$N$_{1+x}$          &  \verb~P6322~    &         110  &    38.09  &        2.3604  &            6  &          184.5  &  \cite{Jacobs:1995_JAC}        \\
 $\epsilon$-Fe$_{3}$N $         $        &                           &  \verb~P312~     &         110  &    38.13  &        2.3580  &            6  &         1000.0  &  \cite{Jack:1952_AC}           \\
 Fe$_{24}$N$_{10}$      &                           &  \verb~P312~     &         220  &    39.07  &        2.3038  &            6  &         1000.0  &  \cite{Jack:1952_AC}           \\
 Fe$_{0.7}$Ga$_{0.3}          $      &  Mg                       &  \verb~P63/mmc~  &         100  &    39.74  &        2.2664  &            6  &          255.2  &  \cite{Schubert:1960_DN}       \\
 FeGa$_{3}                  $        &  FeGa$_{3}$               &  \verb~P42/mnm~  &         202  &    39.78  &        2.2641  &            8  &          272.3  &  \cite{Haussermann:2002_JSSC}  \\
 Fe$_{3}$C$                   $        &  Fe$_{3}$C                &  \verb~Pnma~     &         002  &    39.79  &        2.2635  &            2  &          194.2  &  \cite{Fruchart:1984_JSSC}     \\
 $\epsilon$-Fe$_{3}C          $        &  Fe$_{3}$N$_{1+x}$        &  \verb~P6322~    &         110  &    40.07  &        2.2483  &            6  &          200.9  &  \cite{Lv:2008_CMS}            \\
 Fe$_{3}$C$                   $        &  Fe$_{3}$C                &  \verb~Pnma~     &         201  &    40.62  &        2.2191  &            4  &          167.8  &  \cite{Fruchart:1984_JSSC}     \\
 FeGa$_{3}                  $        &  FeGa$_{3}$               &  \verb~P42/mnm~  &         220  &    40.71  &        2.2142  &            4  &          121.1  &  \cite{Haussermann:2002_JSSC}  \\
 $\epsilon$-Fe$_{2.3}$N $       $        &  NiAs                     &  \verb~P63/mmc~  &         002  &    40.72  &        2.2140  &            2  &          202.6  &  \cite{Kim:994_JJAP}           \\
 Fe$_{24}$N$_{10}$      &                           &  \verb~P312~     &         310  &    40.73  &        2.2134  &            6  &          170.0  &  \cite{Jack:1952_AC}           \\
 Fe$_{24}$N$_{10}$                   &                           &  \verb~P312~     &         130  &    40.73  &        2.2134  &            6  &          170.0  &  \cite{Jack:1952_AC}           \\
 $\zeta$-Fe$_{2}$N $            $        &                           &  \verb~Pbcn~     &         200  &    40.77  &        2.2115  &            2  &          310.0  &  \cite{Hasegawa:2005_JAC}      \\
 $\epsilon$-Fe$_{3}$N $         $        &  Fe$_{3}$N$_{1+x}$        &  \verb~P6322~    &         002  &    40.81  &        2.2094  &            2  &          235.2  &  \cite{Jacobs:1995_JAC}        \\
 $\epsilon$-Fe$_{3}N_{1.39}   $      &  Mn$_{2}$N$_{1-x}$        &  \verb~P6322~    &         002  &    40.82  &        2.2085  &            2  &          209.6  &  \cite{Leineweber:2001_JAC}    \\
 $\epsilon$-Fe$_{2}$N $         $        &                           &  \verb~P-3m1~    &         002  &    40.89  &        2.2050  &            2  &          373.2  &  \cite{Pinsker:1954_DAN}       \\
 $\gamma^{\prime}$-GaFe$_{3}$N            &  CaTiO$_{3}$              &  \verb~Pm-3m~    &         111  &    41.14  &        2.1924  &            8  &         1000.0  &  \cite{Houben:2009_CM}         \\
 (Ga$_{0.5}$Fe$_{0.5}$)Fe$_{3}$N    &  CaTiO$_{3}$              &  \verb~Pm-3m~    &         111  &    41.18  &        2.1905  &            8  &         1000.0  &  \cite{Houben:2009_CM}         \\
 $\gamma^{\prime}$-Fe$_{4}$N $       $      &  CaTiO$_{3}$              &  \verb~Pm-3m~    &         111  &    41.22  &        2.1882  &            8  &         1000.0  &  \cite{Jacobs:1995_JAC}        \\
\hline
 \rowcolor{blue!20} Al$_{2}$O$_{3}$      &  Al$_{2}$O$_{3}$          &  \verb~R-3c:H~   &         006  &    41.68  &        2.1650  &            2  &            1.5  &  \cite{Graham:1960_JPCS}       \\
\hline
 $\epsilon$-Fe                         &  Mg $           $       &  \verb~63/mmc~   &         100  &    42.30  &        2.1348  &            6  &          256.8  &  \cite{Clendenen:1964_JPCS}    \\
 FeGa$_{3}                    $      &  FeGa$_{3}      $     &  \verb~P42/mnm~  &         212  &    42.42  &        2.1292  &           16  &         1000.0  &  \cite{Haussermann:2002_JSSC}  \\
 Fe$_{3}$Ga $                   $      &  AuCu$_{3}      $     &  \verb~Pm-3m~    &         111  &    42.53  &        2.1241  &            8  &         1000.0  &  \cite{Suzuki:1984_MTA}        \\
 $\zeta$-Fe$_{2}$N $              $      &  $              $       &  \verb~Pbcn~     &         102  &    42.68  &        2.1166  &            4  &          500.2  &  \cite{Hasegawa:2005_JAC}      \\
 $\alpha$-Fe$_{16}$N$_{2}         $    &  $              $       &  \verb~I4/mmm~   &         202  &    42.70  &        2.1159  &            8  &         1000.0  &  \cite{Sawada:1994_PRB}        \\
 $\gamma$-Fe                           &  Cu$            $       &  \verb~Fm-3m~    &         111  &    42.83  &        2.1099  &            8  &         1000.0  &  \cite{Straumanis:1969_ZM}     \\
 $\zeta$-Fe$_{2}$N $              $      &  $              $       &  \verb~Pbcn~     &         121  &    42.85  &        2.1087  &            8  &         1000.0  &  \cite{Hasegawa:2005_JAC}      \\
 Fe$_{3}$C$                     $      &  Fe$_{3}$C$       $     &  \verb~Pnma~     &         211  &    42.87  &        2.1078  &            8  &          576.0  &  \cite{Fruchart:1984_JSSC}     \\
 Fe$_{0.7}$Ga$_{0.3}            $    &  Mg $           $       &  \verb~P63/mmc~  &         002  &    42.89  &        2.1070  &            2  &          267.4  &  \cite{Schubert:1960_DN}       \\
 $\gamma$-FeN$_{0.0950}         $      &  $\gamma$-FeN$_{x}$         &  \verb~Fm-3m~    &         111  &    42.93  &        2.1050  &            8  &         1000.0  &  \cite{Jack:1951_PRSLA}        \\
 $\epsilon$-Fe$_{2}$N $           $      &  $              $       &  \verb~P-3m1~    &         011  &    42.94  &        2.1044  &            6  &         1000.0  &  \cite{Pinsker:1954_DAN}       \\
 $\epsilon$-Fe$_{2}$N $           $      &  $              $       &  \verb~P-3m1~    &         101  &    42.94  &        2.1044  &            6  &          819.2  &  \cite{Pinsker:1954_DAN}       \\
 $\epsilon$-Fe$_{3}$N$_{1.39}     $    &  Mn$_{2}$N$_{1-x} $   &  \verb~P6322~    &         111  &    42.95  &        2.1039  &           12  &         1000.0  &  \cite{Leineweber:2001_JAC}    \\
 $\epsilon$-Fe$_{3}$C $           $      &  Fe$_{3}$N$_{1+x} $   &  \verb~P6322~    &         002  &    43.00  &        2.1018  &            2  &          247.7  &  \cite{Lv:2008_CMS}            \\
 Fe$_{3}$O$_{4}                 $    &  Al$_{2}$MgO$_{4} $   &  \verb~Fd-3mZ~   &         400  &    43.07  &        2.0985  &            6  &          201.3  &  \cite{Fleet:1981_ACB}         \\
 Fe$_{24}$N$_{10}$                   &  $              $       &  \verb~P312~     &         102  &    43.13  &        2.0958  &            6  &          209.7  &  \cite{Jack:1952_AC}           \\
 $\alpha$-FeN$_{0.0950}         $      &  $\alpha$-FeN$_{x}$         &  \verb~I4/mmm~   &         101  &    43.15  &        2.0950  &            8  &         1000.0  &  \cite{Jack:1951_PRSLA}        \\
 $\gamma$-Fe                           &  Cu$            $       &  \verb~Fm-3m~    &         111  &    43.38  &        2.0842  &            8  &         1000.0  &  \cite{Westgren:1921_ZPC}      \\
 $\epsilon$-Fe$_{3}$N $           $      &  Fe$_{3}$N$_{1+x} $   &  \verb~P6322~    &         111  &    43.43  &        2.0820  &           12  &         1000.0  &  \cite{Jacobs:1995_JAC}        \\
 $\alpha$-Fe                           &  W $            $       &  \verb~Im-3m~    &         110  &    43.63  &        2.0729  &           12  &         1000.0  &  \cite{Basinski:1955_PRSLA}    \\
 Fe$_{3}$C$                     $      &  Fe$_{3}$C$       $     &  \verb~Pnma~     &         102  &    43.73  &        2.0684  &            4  &          565.9  &  \cite{Fruchart:1984_JSSC}     \\
 Fe$_{0.8}$Ga$_{0.2}            $    &  W $            $       &  \verb~Im-3m~    &         110  &    44.02  &        2.0556  &           12  &         1000.0  &  \cite{Buschow:1983_JMMM}      \\
 Fe$_{3}$Ga $                   $      &  $BiF_{3}       $     &  \verb~Fm-3m~    &         220  &    44.07  &        2.0531  &           12  &         1000.0  &  \cite{Nishino:1991_SMM}       \\
 $\alpha$-Fe                           &  W $            $       &  \verb~Im-3m~    &         110  &    44.35  &        2.0407  &           12  &         1000.0  &  \cite{Swanson:1955_NBS}       \\
 Fe$_{3}$C$                     $      &  Fe$_{3}$C$       $     &  \verb~Pnma~     &         220  &    44.56  &        2.0315  &            4  &          526.2  &  \cite{Fruchart:1984_JSSC}     \\
 $\alpha$-Fe                           &  W $            $       &  \verb~Im-3m~    &         110  &    44.68  &        2.0267  &           12  &         1000.0  &  \cite{Straumanis:1969_ZM}     \\
 $\epsilon$-Fe                         &  Mg $           $       &  \verb~P63/mmc~  &         002  &    44.72  &        2.0250  &            2  &          285.4  &  \cite{Clendenen:1964_JPCS}    \\
 $\alpha$-Fe$_{16}$N$_{2}         $    &  $              $       &  \verb~I4/mmm~   &         220  &    44.78  &        2.0223  &            4  &          510.2  &  \cite{Sawada:1994_PRB}        \\
 $\alpha$-FeN$_{0.0950}         $      &  $\alpha$-FeN$_{x}$         &  \verb~I4/mmm~   &         110  &    44.94  &        2.0153  &            4  &          473.3  &  \cite{Jack:1951_PRSLA}        \\
 Fe$_{3}$C$                     $      &  Fe$_{3}$C$       $     &  \verb~Pnma~     &         031  &    45.00  &        2.0127  &            4  &         1000.0  &  \cite{Fruchart:1984_JSSC}     \\
 Fe$_{0.7}$Ga$_{0.3}            $    &  Mg $           $       &  \verb~P63/mmc~  &         101  &    45.40  &        1.9960  &           12  &         1000.0  &  \cite{Schubert:1960_DN}       \\
 $\epsilon$-Fe$_{3}$C $           $      &  Fe$_{3}$N$_{1+x} $   &  \verb~P6322~    &         111  &    45.73  &        1.9825  &           12  &         1000.0  &  \cite{Lv:2008_CMS}            \\
 $\gamma$-Fe                           &  Cu$            $       &  \verb~Fm-3m~    &         111  &    45.78  &        1.9803  &            8  &         1000.0  &  \cite{Haglund:1993_PRB}       \\
 FeGa$_{3}                    $      &  FeGa$_{3}      $     &  \verb~P42/mnm~  &         310  &    45.78  &        1.9805  &            8  &          561.4  &  \cite{Haussermann:2002_JSSC}  \\
 Fe$_{3}$C$                     $      &  Fe$_{3}$C$       $     &  \verb~Pnma~     &         112  &    45.85  &        1.9774  &            8  &          487.1  &  \cite{Fruchart:1984_JSSC}     \\
 Fe$_{24}$N$_{10}$                   &  $              $       &  \verb~P312~     &         112  &    46.17  &        1.9646  &            6  &          199.8  &  \cite{Jack:1952_AC}           \\
 Fe$_{24}$N$_{10}$                   &  $              $       &  \verb~P312~     &        11$\bar{2}$  &    46.17  &        1.9646  &            6  &          199.8  &  \cite{Jack:1952_AC}           \\
 $\epsilon$-Fe$_{3}$N $           $      &  $              $       &  \verb~P312~     &         012  &    46.92  &        1.9348  &            6  &          168.8  &  \cite{Jack:1952_AC}           \\
 $\gamma^{\prime}$-GaFe$_{3}$N            &  CaTiO$_{3}     $     &  \verb~Pm-3m~    &         200  &    47.87  &        1.8987  &            6  &          557.0  &  \cite{Houben:2009_CM}         \\
 (Ga$_{0.5}$Fe$_{0.5}$)Fe$_{3}$N     &  CaTiO$_{3}     $     &  \verb~Pm-3m~    &         200  &    47.91  &        1.8971  &            6  &          558.2  &  \cite{Houben:2009_CM}         \\
 $\gamma^{\prime}$-Fe$_{4}$N $         $    &  CaTiO$_{3}     $     &  \verb~Pm-3m~    &         200  &    47.97  &        1.8950  &            6  &          569.3  &  \cite{Jacobs:1995_JAC}        \\
 $\epsilon$-Fe                         &  Mg $           $       &  \verb~P63/mmc~  &         101  &    48.18  &        1.8885  &           12  &         1000.0  &  \cite{Clendenen:1964_JPCS}    \\
 $\epsilon$-Fe$_{2}$N $           $      &  $              $       &  \verb~P312~     &         021  &    48.47  &        1.8767  &            6  &          215.5  &  \cite{Jack:1952_AC}           \\
 Fe$_{3}$C$                     $      &  Fe$_{3}$C$       $     &  \verb~Pnma~     &         131  &    48.60  &        1.8718  &            8  &          332.7  &  \cite{Fruchart:1984_JSSC}     \\
 Fe$_{3}$C$                     $      &  Fe$_{3}$C$       $     &  \verb~Pnma~     &         221  &    49.11  &        1.8535  &            8  &          409.1  &  \cite{Fruchart:1984_JSSC}     \\
 Fe$_{2}$O$_{3}                 $    &  Al$_{2}$O$_{3}$   &  \verb~R-3c:H~   &         024  &    49.42  &        1.8428  &            6  &          359.4  &  \cite{Blake:1966_AM}          \\
 Fe$_{3}$Ga $                   $      &  AuCu$_{3}      $     &  \verb~Pm-3m~    &         200  &    49.51  &        1.8395  &            6  &          452.6  &  \cite{Suzuki:1984_MTA}        \\
 $\epsilon$-Fe$_{3}$N $           $      &  $              $       &  \verb~P312~     &         021  &    49.16  &        1.8519  &            6  &          145.1  &  \cite{Jack:1952_AC}           \\
 $\gamma$-Fe                           &  Cu$            $       &  \verb~Fm-3m~    &         200  &    49.87  &        1.8272  &            6  &          448.5  &  \cite{Straumanis:1969_ZM}     \\
 $\gamma$-FeN$_{0.0950}         $      &  $\gamma$-FeN$_{x}$         &  \verb~Fm-3m~    &         200  &    49.99  &        1.8230  &            6  &          484.7  &  \cite{Jack:1951_PRSLA}        \\
 $\gamma$-Fe                           &  Cu$            $       &  \verb~Fm-3m~    &         200  &    50.52  &        1.8050  &            6  &          446.7  &  \cite{Westgren:1921_ZPC}      \\
 Fe$_{3}$C$                     $      &  Fe$_{3}$C$       $     &  \verb~Pnma~     &         122  &    51.82  &        1.7629  &            8  &          170.7  &  \cite{Fruchart:1984_JSSC}     \\
 $\gamma$-Fe                           &  Cu$            $       &  \verb~Fm-3m~    &         200  &    53.38  &        1.7150  &            6  &          439.3  &  \cite{Haglund:1993_PRB}       \\
 Fe$_{2}$O$_{3}                 $    &  Al$_{2}$O$_{3}$      &  \verb~R-3c:H~   &         116  &    54.00  &        1.6966  &            6  &          213.7  &  \cite{Blake:1966_AM}          \\
 Fe$_{2}$O$_{3}                 $    &  Al$_{2}$O$_{3}$      &  \verb~R-3c:H~   &        11$\bar{6}$  &    54.00  &        1.6966  &            6  &          213.7  &  \cite{Blake:1966_AM}          \\
 Fe$_{3}$C$                     $      &  Fe$_{3}$C$       $     &  \verb~Pnma~     &         230  &    54.42  &        1.6847  &            4  &          116.1  &  \cite{Fruchart:1984_JSSC}     \\
 $\gamma$-Fe                           &  Cu$            $       &  \verb~Fm-3m~    &         111  &    55.60  &        1.6517  &            8  &         1000.0  &  \cite{Owen:1921_PM}           \\
 FeGa$_{3}                    $      &  FeGa$_{3}      $     &  \verb~P42/mnm~  &         004  &    56.08  &        1.6386  &            2  &          160.8  &  \cite{Haussermann:2002_JSSC}  \\
 $\zeta$-Fe$_{2}$N $              $      &  $              $       &  \verb~bcn~      &         221  &    56.55  &        1.6260  &            8  &          176.6  &  \cite{Hasegawa:2005_JAC}      \\
 $\epsilon$-Fe$_{2.3}$N $         $      &  NiAs $         $       &  \verb~P63/mmc~  &         102  &    56.65  &        1.6235  &           12  &          764.3  &  \cite{Kim:994_JJAP}           \\
 $\epsilon$-Fe$_{3}N_{1.39}     $    &  Mn$_{2}$N$_{1-x} $   &  \verb~P6322~    &         112  &    56.67  &        1.6229  &           12  &          189.7  &  \cite{Leineweber:2001_JAC}    \\
 Fe$_{3}$O$_{4}                 $    &  Al$_{2}$MgO$_{4} $   &  \verb~Fd-3mZ~   &         511  &    56.96  &        1.6154  &           24  &          235.6  &  \cite{Fleet:1981_ACB}         \\
 $\epsilon$-Fe$_{3}$N $           $      &  Fe$_{3}$N$_{1+x} $   &  \verb~P6322~    &         112  &    57.05  &        1.6130  &           12  &          166.7  &  \cite{Jacobs:1995_JAC}        \\
 Fe$_{3}$C$                     $      &  Fe$_{3}$C$       $     &  \verb~Pnma~     &         301  &    57.98  &        1.5893  &            4  &          151.0  &  \cite{Fruchart:1984_JSSC}     \\
 Fe$_{0.7}$Ga$_{0.3}$                &  Mg $           $       &  \verb~P63/mmc~  &         102  &    59.89  &        1.5431  &           12  &          126.5  &  \cite{Schubert:1960_DN}       \\
 $\epsilon$-Fe$_{3}$C $           $      &  Fe$_{3}$N$_{1+x} $   &  \verb~P6322~    &         112  &    60.22  &        1.5354  &           12  &          150.1  &  \cite{Lv:2008_CMS}            \\
 Fe$_{2}$O$_{3}                 $    &  Al$_{2}$O$_{3}$      &  \verb~R-3c:H~   &        12$\bar{4}$  &    62.38  &        1.4873  &            6  &          141.2  &  \cite{Blake:1966_AM}          \\
 Fe$_{2}$O$_{3}                 $    &  Al$_{2}$O$_{3}$      &  \verb~R-3c:H~   &         214  &    62.38  &        1.4873  &            6  &          141.2  &  \cite{Blake:1966_AM}          \\
 Fe$_{3}$O$_{4}                 $    &  Al$_{2}$MgO$_{4} $   &  \verb~Fd-3mZ~   &         440  &    62.54  &        1.4839  &           12  &          388.9  &  \cite{Fleet:1981_ACB}         \\
 $\epsilon$-Fe                         &  Mg $           $       &  \verb~P63/mmc~  &         102  &    63.24  &        1.4692  &           12  &          123.7  &  \cite{Clendenen:1964_JPCS}    \\
 $\alpha$-Fe                           &  W $            $       &  \verb~Im-3m~    &         200  &    63.41  &        1.4657  &            6  &          131.9  &  \cite{Basinski:1955_PRSLA}    \\
 Fe$_{2}$O$_{3}                 $    &  Al$_{2}$O$_{3}$      &  \verb~R-3c:H~   &         300  &    63.96  &        1.4543  &            6  &          270.2  &  \cite{Blake:1966_AM}          \\
 Fe$_{0.8}$Ga$_{0.2}            $    &  W $            $       &  \verb~Im-3m~    &         200  &    64.00  &        1.4535  &            6  &          133.1  &  \cite{Buschow:1983_JMMM}      \\
 Fe$_{3}$Ga $                   $      &  $BiF_{3}       $     &  \verb~Fm-3m~    &         400  &    64.09  &        1.4518  &            6  &          133.5  &  \cite{Nishino:1991_SMM}       \\
 $\alpha$-Fe                           &  W $            $       &  \verb~Im-3m~    &         200  &    64.53  &        1.4430  &            6  &          130.3  &  \cite{Swanson:1955_NBS}       \\
 $\alpha$-Fe                           &  W $            $       &  \verb~Im-3m~    &         200  &    65.03  &        1.4331  &            6  &          129.7  &  \cite{Straumanis:1969_ZM}     \\
 $\gamma$-Fe                           &  Cu$            $       &  \verb~Fm-3m~    &         200  &    65.16  &        1.4304  &            6  &          414.5  &  \cite{Owen:1921_PM}           \\
 $\alpha$-Fe$_{16}$N$_{2}         $    &  $              $       &  \verb~I4/mmm~   &         400  &    65.18  &        1.4300  &            4  &          130.2  &  \cite{Sawada:1994_PRB}        \\
 $\alpha$-FeN$_{0.0950}         $      &  $\alpha$-FeN$_{x}$         &  \verb~I4/mmm~   &         200  &    65.44  &        1.4250  &            4  &          121.6  &  \cite{Jack:1951_PRSLA}        \\
 $\zeta$-Fe$_{2}$N $              $      &  $              $       &  \verb~Pbcn~     &         023  &    67.34  &        1.3894  &            4  &          155.8  &  \cite{Hasegawa:2005_JAC}      \\
 $\epsilon$-Fe$_{2}$N $           $      &  $              $       &  \verb~P-3m1~    &         110  &    67.72  &        1.3825  &            6  &          309.9  &  \cite{Pinsker:1954_DAN}       \\
 $\epsilon$-Fe$_{3}$N $           $      &  $              $       &  \verb~P312~     &         300  &    67.72  &        1.3825  &            6  &          287.6  &  \cite{Pinsker:1954_DAN}       \\
 $\epsilon$-Fe$_{2}$N $           $      &  $              $       &  \verb~P312~     &         300  &    67.75  &        1.3819  &            6  &          175.9  &  \cite{Jack:1952_AC}           \\
 $\epsilon$-Fe$_{3}N_{1.39}     $    &  Mn$_{2}$N$_{1-x} $   &  \verb~P6322~    &         300  &    67.78  &        1.3815  &            6  &          176.6  &  \cite{Leineweber:2001_JAC}    \\
 $\epsilon$-Fe$_{2.3}$N $         $      &  NiAs $         $       &  \verb~P63/mmc~  &         110  &    67.94  &        1.3785  &            6  &          223.8  &  \cite{Kim:994_JJAP}           \\
 FeGa$_{3}                    $      &  FeGa$_{3}      $     &  \verb~P42/mnm~  &         412  &    67.96  &        1.3781  &           16  &          130.0  &  \cite{Haussermann:2002_JSSC}  \\
 $\epsilon$-Fe$_{3}$N $           $      &  Fe$_{3}$N$_{1+x} $   &  \verb~P6322~    &         300  &    68.83  &        1.3628  &            6  &          154.6  &  \cite{Jacobs:1995_JAC}        \\
 $\epsilon$-Fe$_{3}$N $           $      &  $              $       &  \verb~P312~     &         300  &    68.92  &        1.3614  &            6  &          282.8  &  \cite{Jack:1952_AC}           \\
 FeGa$_{3}                    $      &  FeGa$_{3}      $     &  \verb~P42/mnm~  &         332  &    69.82  &        1.3459  &            8  &          110.1  &  \cite{Haussermann:2002_JSSC}  \\
 $\gamma^{\prime}$-GaFe$_{3}$N            &  CaTiO$_{3}     $     &  \verb~Pm-3m~    &         220  &    70.02  &        1.3426  &           12  &          277.6  &  \cite{Houben:2009_CM}         \\
 (Ga$_{0.5}$Fe$_{0.5}$)Fe$_{3}$N     &  CaTiO$_{3}     $     &  \verb~Pm-3m~    &         220  &    70.09  &        1.3414  &           12  &          275.8  &  \cite{Houben:2009_CM}         \\
 $\gamma^{\prime}$-Fe$_{4}$N $         $    &  CaTiO$_{3}     $     &  \verb~Pm-3m~    &         220  &    70.18  &        1.3400  &           12  &          295.8  &  \cite{Jacobs:1995_JAC}        \\
 Fe$_{24}$N$_{10}$                   &  $              $       &  \verb~P312~     &         600  &    70.78  &        1.3301  &            6  &          297.9  &  \cite{Jack:1952_AC}           \\
 Fe$_{3}$C$                     $      &  Fe$_{3}$C$       $     &  \verb~Pnma~     &         123  &    70.81  &        1.3295  &            8  &          153.0  &  \cite{Fruchart:1984_JSSC}     \\
 Fe$_{0.7}$Ga$_{0.3}$                &  Mg $           $       &  \verb~P63/mmc~  &         110  &    72.13  &        1.3085  &            6  &          127.2  &  \cite{Schubert:1960_DN}       \\
 Fe$_{3}$Ga $                   $      &  AuCu$_{3}      $     &  \verb~Pm-3m~    &         220  &    72.63  &        1.3007  &           12  &          224.5  &  \cite{Suzuki:1984_MTA}        \\
 $\epsilon$-Fe$_{3}$C $           $      &  Fe$_{3}$N$_{1+x} $   &  \verb~P6322~    &         300  &    72.80  &        1.2980  &            6  &          129.7  &  \cite{Lv:2008_CMS}            \\
\hline
 \rowcolor{red!20}  GaN                  &  ZnS(2H)                  &  \verb~P63mc~    &         004  &    72.90  &        1.2965  &            2  &           25.9  &  \cite{Paszkowicz:2004_JAC}    \\
\hline
 $\gamma$-Fe                           &  Cu$             $      &  \verb~Fm-3m~    &         220  &    73.19  &        1.2920  &           12  &          217.1  &  \cite{Straumanis:1969_ZM}     \\
 $\gamma$-FeN$_{0.0950}        $       &  $\gamma$-FeN$_{x} $    &  \verb~Fm-3m~    &         220  &    73.39  &        1.2891  &           12  &          233.3  &  \cite{Jack:1951_PRSLA}        \\
 $\gamma$-Fe                           &  Cu$             $      &  \verb~Fm-3m~    &         220  &    74.24  &        1.2763  &           12  &          214.9  &  \cite{Westgren:1921_ZPC}      \\
 FeGa$_{3}                   $       &  FeGa$_{3}       $    &  \verb~P42/mnm~  &         314  &    75.20  &        1.2625  &           16  &          227.9  &  \cite{Haussermann:2002_JSSC}  \\
 $\zeta$-Fe$_{2}$N $             $       &  $               $      &  \verb~Pbcn~     &         321  &    75.65  &        1.2561  &            8  &          128.5  &  \cite{Hasegawa:2005_JAC}      \\
 $\epsilon$-Fe$_{3}N_{1.39}    $     &  Mn$_{2}$N$_{1-x}  $  &  \verb~P6322~    &         113  &    75.80  &        1.2540  &           12  &          157.7  &  \cite{Leineweber:2001_JAC}    \\
 $\epsilon$-Fe$_{3}$N $          $       &  Fe$_{3}$N$_{1+x}  $  &  \verb~P6322~    &         113  &    76.11  &        1.2496  &           12  &          151.7  &  \cite{Jacobs:1995_JAC}        \\
 $\epsilon$-Fe$_{3}$N $          $       &  $               $      &  \verb~P312~     &         113  &    76.50  &        1.2442  &            6  &          140.7  &  \cite{Jack:1952_AC}           \\
 $\epsilon$-Fe$_{3}$N $          $       &  $               $      &  \verb~P312~     &        11$\bar{3}$  &    76.50  &        1.2442  &            6  &          140.7  &  \cite{Jack:1952_AC}           \\
 $\alpha$-Fe$_{16}$N$_{2}        $     &  $               $      &  \verb~I4/mmm~   &         224  &    76.71  &        1.2414  &            8  &          108.2  &  \cite{Sawada:1994_PRB}        \\
 $\epsilon$-Fe                         &  Mg $            $      &  \verb~P63/mmc~  &         110  &    77.36  &        1.2325  &            6  &          116.9  &  \cite{Clendenen:1964_JPCS}    \\
 $\alpha$-FeN$_{0.0950}        $       &  $\alpha$-FeN$_{x} $    &  \verb~I4/mmm~   &         112  &    77.84  &        1.2261  &            8  &          127.7  &  \cite{Jack:1951_PRSLA}        \\
 Fe$_{3}$C$                    $       &  Fe$_{3}$C$        $    &  \verb~Pnma~     &         401  &    77.89  &        1.2255  &            4  &          145.1  &  \cite{Fruchart:1984_JSSC}     \\
 Fe$_{3}$C$                    $       &  Fe$_{3}$C$        $    &  \verb~Pnma~     &         133  &    78.58  &        1.2165  &            8  &          191.3  &  \cite{Fruchart:1984_JSSC}     \\
 $\gamma$-Fe                           &  Cu$             $      &  \verb~Fm-3m~    &         220  &    78.87  &        1.2127  &           12  &          207.0  &  \cite{Haglund:1993_PRB}       \\
 $\alpha$-Fe                           &  W $             $      &  \verb~Im-3m~    &         211  &    80.13  &        1.1968  &           24  &          222.6  &  \cite{Basinski:1955_PRSLA}    \\
 Fe$_{0.7}$Ga$_{0.3}$                &  Mg $            $      &  \verb~P63/mmc~  &         103  &    80.35  &        1.1939  &           12  &          130.2  &  \cite{Schubert:1960_DN}       \\
 $\epsilon$-Fe$_{3}$C $          $       &  Fe$_{3}$N$_{1+x}  $  &  \verb~P6322~    &         113  &    80.74  &        1.1892  &           12  &          126.8  &  \cite{Lv:2008_CMS}            \\
 Fe$_{3}$Ga $                  $       &  $BiF_{3}        $    &  \verb~Fm-3m~    &         422  &    81.06  &        1.1853  &           24  &          228.2  &  \cite{Nishino:1991_SMM}       \\
 $\alpha$-Fe$_{16}$N$_{2}        $     &  $               $      &  \verb~I4/mmm~   &         422  &    81.10  &        1.1848  &           16  &          198.0  &  \cite{Sawada:1994_PRB}        \\
 $\alpha$-Fe                           &  W $             $      &  \verb~Im-3m~    &         211  &    81.65  &        1.1782  &           24  &          220.5  &  \cite{Swanson:1955_NBS}       \\
 $\alpha$-FeN$_{0.0950}        $       &  $\alpha$-FeN$_{x} $    &  \verb~I4/mmm~   &         211  &    81.65  &        1.1783  &           16  &          203.0  &  \cite{Jack:1951_PRSLA}        \\
 $\epsilon$-Fe$_{3}N_{1.39}    $     &  Mn$_{2}$N$_{1-x}  $  &  \verb~P6322~    &         302  &    82.25  &        1.1712  &           12  &          140.6  &  \cite{Leineweber:2001_JAC}    \\
 $\epsilon$-Fe$_{2.3}$N $        $       &  NiAs $          $      &  \verb~P63/mmc~  &         112  &    82.33  &        1.1702  &           12  &          110.8  &  \cite{Kim:994_JJAP}           \\
 $\alpha$-Fe                           &  W $             $      &  \verb~Im-3m~    &         211  &    82.34  &        1.1701  &           24  &          219.7  &  \cite{Straumanis:1969_ZM}     \\
 Fe$_{3}$C$                    $       &  Fe$_{3}$C$        $    &  \verb~Pnma~     &         332  &    83.02  &        1.1622  &            8  &          200.4  &  \cite{Fruchart:1984_JSSC}     \\
 $\epsilon$-Fe$_{3}$N $          $       &  Fe$_{3}$N$_{1+x}  $  &  \verb~P6322~    &         302  &    83.23  &        1.1599  &           12  &          140.6  &  \cite{Jacobs:1995_JAC}        \\
 $\epsilon$-Fe$_{3}N_{1.39}    $     &  Mn$_{2}$N$_{1-x}  $  &  \verb~P6322~    &         221  &    83.68  &        1.1548  &           12  &          118.8  &  \cite{Leineweber:2001_JAC}    \\
 Fe$_{3}$C$                    $       &  Fe$_{3}$C$        $    &  \verb~Pnma~     &         251  &    83.91  &        1.1522  &            8  &          110.2  &  \cite{Fruchart:1984_JSSC}     \\
 $\gamma^{\prime}$-GaFe$_{3}$N            &  CaTiO$_{3}      $    &  \verb~Pm-3m~    &         311  &    84.56  &        1.1450  &           24  &          244.3  &  \cite{Houben:2009_CM}         \\
 $\gamma^{\prime}$-Fe$_{4}$N $        $     &  CaTiO$_{3}      $    &  \verb~Pm-3m~    &         311  &    84.76  &        1.1427  &           24  &          266.4  &  \cite{Jacobs:1995_JAC}        \\
 $\epsilon$-Fe                         &  Mg $            $      &  \verb~P63/mmc~  &         103  &    84.92  &        1.1410  &           12  &          128.1  &  \cite{Clendenen:1964_JPCS}    \\
 $\epsilon$-Fe$_{3}$N $          $       &  Fe$_{3}$N$_{1+x}  $  &  \verb~P6322~    &         221  &    84.99  &        1.1403  &           12  &          108.2  &  \cite{Jacobs:1995_JAC}        \\
 Fe$_{3}$C$                    $       &  Fe$_{3}$C$        $    &  \verb~Pnma~     &         303  &    86.16  &        1.1278  &            4  &          123.2  &  \cite{Fruchart:1984_JSSC}     \\
 Fe$_{0.7}$Ga$_{0.3}$                &  Mg $            $      &  \verb~P63/mmc~  &         112  &    87.73  &        1.1116  &           12  &          131.6  &  \cite{Schubert:1960_DN}       \\
 Fe$_{3}$Ga $                  $       &  AuCu$_{3}       $    &  \verb~Pm-3m~    &         311  &    87.97  &        1.1093  &           24  &          236.0  &  \cite{Suzuki:1984_MTA}        \\
 Fe$_{3}$C$                    $       &  Fe$_{3}$C$        $    &  \verb~Pnma~     &         430  &    88.13  &        1.1076  &            4  &          128.8  &  \cite{Fruchart:1984_JSSC}     \\
 $\epsilon$-Fe$_{3}$C $          $       &  Fe$_{3}$N$_{1+x}  $  &  \verb~P6322~    &         302  &    88.45  &        1.1044  &           12  &          110.6  &  \cite{Lv:2008_CMS}            \\
 $\gamma$-Fe                           &  Cu$             $      &  \verb~Fm-3m~    &         311  &    88.71  &        1.1018  &           24  &          227.5  &  \cite{Straumanis:1969_ZM}     \\
 $\gamma$-FeN$_{0.0950}        $       &  $\gamma$-FeN$_{x} $    &  \verb~Fm-3m~    &         311  &    88.97  &        1.0993  &           24  &          229.3  &  \cite{Jack:1951_PRSLA}        \\
 FeGa$_{3}                   $       &  FeGa$_{3}       $    &  \verb~P42/mnm~  &         522  &    89.30  &        1.0960  &           16  &          149.5  &  \cite{Haussermann:2002_JSSC}  \\
 $\gamma$-Fe                           &  Cu$             $      &  \verb~Fm-3m~    &         311  &    90.09  &        1.0885  &           24  &          227.1  &  \cite{Westgren:1921_ZPC}      \\
 $\epsilon$-Fe                         &  Mg $            $      &  \verb~P63/mmc~  &         112  &    94.05  &        1.0528  &           12  &          128.1  &  \cite{Clendenen:1964_JPCS}    \\
 $\epsilon$-Fe$_{2.3}$N                &  NiAs $          $      &  \verb~P63/mmc~  &         202  &    94.29  &        1.0508  &           12  &          127.9  &  \cite{Kim:994_JJAP}           \\
 $\gamma$-Fe                           &  Cu$             $      &  \verb~Fm-3m~    &         311  &    96.29  &        1.0342  &           24  &          230.0  &  \cite{Haglund:1993_PRB}       \\
 $\gamma$-Fe                           &  Cu$             $      &  \verb~Fm-3m~    &         220  &    99.20  &        1.0114  &           12  &          213.9  &  \cite{Owen:1921_PM}           \\
\hline
\end{longtable}

%

\end{document}